\documentclass[twocolumn,prb,amsmath,amssymb,showpacs]{revtex4}

\usepackage{graphicx}

\begin{document}
\title{Small-angle scattering in a marginal Fermi-liquid}
\author{E. C. Carter}
\author{A. J. Schofield}
\affiliation{%
Theoretical Physics Group, School of Physics and Astronomy, University
of Birmingham, Edgbaston, Birmingham B15 2TT, United Kingdom
}%

\date{\today}

\begin{abstract}
We study the magnetotransport properties of a model of small-angle
scattering in a marginal Fermi liquid. Such a model has been proposed
by Varma and Abrahams [Phys. Rev. Lett. 86, 4652 (2001)] to account
for the anomalous temperature dependence of in--plane magnetotransport
properties of the high-$T_c$ cuprates. We study the resistivity, Hall
angle and magnetoresistance using both analytical and numerical
techniques. We find that small-angle scattering only generates a new
temperature dependence for the Hall angle near particle-hole symmetric
Fermi surfaces where the conventional Hall term vanishes. The
magnetoresistance always shows Kohler's rule behavior.
\end{abstract}

\maketitle


The unusual magnetotransport properties in the normal state of the
superconducting cuprates pose a major challenge to existing models of
metallic behavior.  Experiments on near optimally-doped materials
have shown a linear-$T$ dependence of the resistivity, $\rho$,
indicative of a marginal Fermi-liquid scattering
rate~\cite{varma1}. However, the scattering rate of the Hall component
of an electric current exhibits an entirely different temperature
dependence from that of the resistivity~\cite{chien1}. Moreover,
Kohler's rule for the magnetoresistance is also
violated~\cite{harris1,malinowski1}.

An influential hypothesis was proposed by Anderson~\cite{anderson1} to
explain these results. He argued that two independent scattering rates
are present in the cuprates which govern the resistivity and
Hall-current scattering respectively by appearing multiplicatively in
the Hall conductivity. This ``two-lifetime'' scenario has often been
used to interpret experimental data. Within this picture, for example,
the deviations of Kohler's rule in the magnetoresistance are accounted
for by a ``modified Kohler's rule'' whereby the lifetime controlling
the Hall angle is the one that also governs the
magnetoresistance~\cite{harris1}.  There have been many attempts to
give a more microscopic explanation of the ``two-lifetime''
phenomenology, either by invoking a Fermi-liquid picture with
anisotropic scattering effects (for example, hot or cold spots on the
Fermi surface\cite{stojkovic1,ioffe} or skew
scattering\cite{kotliar1}), or going beyond a Fermi-liquid
description~\cite{coleman1,lee1} inspired by Anderson's original
motivation.

More recently Varma and Abrahams (VA)~\cite{va,vaerratum} have
proposed that there is only a single temperature dependent scattering
rate in the cuprates but this rate appears in magnetotransport
responses in an unconventional fashion. In particular, the Hall
angle is given by the square of the marginal scattering lifetime. 
This idea has received experimental support from very recent 
infrared optical Hall angle studies\cite{grayson1}. The optical 
data fits best to a square Lorentzian in exactly the manner that Varma
and Abrahams would predict.

Given this experimental support for VA's hypothesis, we re-examine the
microscopic basis they used to derive their form for the Hall
conductivity. They considered how transport in a marginal Fermi-liquid
is affected by small-angle impurity scattering anisotropically
distributed around the Fermi surface. They make an expansion in the
small scattering angle and argue that in the Hall conductivity the
first term in the expansion dominates the conventional (zeroth order)
term. In the longitudinal conductivity the conventional term always
dominates. 

In this paper, we address a number of key questions raised by their
work, for when a perturbative correction dominates the zeroth order
term it is usually an indication that the expansion is breaking
down. Thus, we must establish first: under what conditions does the
first correction dominates the zeroth order term? Secondly, we
must check whether the expansion remains controlled under such
conditions. Finally we consider the magnetoresistance since, as we
will show, it is necessary for higher order terms to dominate the
expansion in the magnetoconductance if the deviations from Kohler's
rule seen in experiment are to be accounted for within this model.
Our approach combines an analytic expansion in the small-angle
scattering parameter together with an extensive numerical
investigation including anisotropy in the Fermi velocity and the small
angle of scattering.

Our main findings are as follows. We find that for both the
longitudinal conductivity and the Hall conductivity the corrections
due to small-angle scattering are generically small compared to the
leading term~\cite{hlubina}. However by tuning the Fermi surface close to
particle-hole symmetry (zero average curvature) the conventional term
can of course be tuned to zero. So under these conditions the Hall
conductance is dominated by the leading correction in a controlled
fashion leading to the form suggested originally by Varma and
Abrahams. It is here that we emphasize the importance of the
magnetoresistance as a key test for any microscopic theory (see for
example Ref.~\onlinecite{sandeman1}). We calculate the
magnetoresistance directly and find that no new temperature dependence
generated even at particle-hole symmetry.

We begin by describing our model.  Varma and Abrahams' suggestion of
anisotropic small-angle scattering processes in a marginal
Fermi-liquid was inspired by angle-resolved photoemission spectroscopy
(ARPES) measurements which indicate that the scattering rate of
electrons has the following form:
\begin{equation}
\label{mt-arpesgamma}
\Gamma \sim \tau_M^{-1} + \tau_i^{-1} \cos^2 2\theta\;,
\end{equation}
where there is an isotropic temperature dependent contribution,
$\tau_M^{-1} \sim A + B T$ (the marginal Fermi-liquid inverse
lifetime) and also an anisotropic temperature-independent contribution
(proportional to $\tau_i^{-1}$). The idea of Ref. \onlinecite{va} is
that $\tau_i^{-1}$ is an outcome of small-angle scattering processes
(possibly from out-of-plane impurities).

Here, we introduce (but do not restrict ourselves to) an explicit form
of the scattering rate which satisfies these criteria:
\begin{equation}
\label{mt-thetathetaeqn}
\tau^{-1}(\theta,\theta^\prime) =
\tau_M^{-1}+
\tau_i^{-1} |\cos 2\theta||\cos 2\theta^\prime|
e^{\frac{-{(\theta-\theta^\prime)}^2}{2\theta_c^2}}/\theta_c,
\end{equation}
where $\theta,\theta^\prime$ measure distance round the Fermi surface.
\footnote{The moduli are unimportant as the narrow Gaussian
only allows contributions where $|\theta-\theta^\prime|\lesssim\theta_c$.}

The linearized Boltzmann transport equation\cite{kotliar1,va} is
\begin{multline}
\label{adef-eqn}
\sum_{{\mathbf k}^\prime}
\left[\left(
1/\tau({\mathbf k}) + \frac{e}{\hbar} {\mathbf v}_{\mathbf k} \times \mathbf
B \cdot \nabla_{\mathbf k} \right) 
\delta_{\mathbf k,{\mathbf k}^\prime}
- C(\mathbf k,{\mathbf k}^\prime)
\right]
g({\mathbf k}^\prime) \\
\equiv
\sum_{{\mathbf k}^\prime} A({\mathbf k},{\mathbf k}^\prime) g({\mathbf
k}^\prime) 
=
e \mathbf E \cdot {\mathbf v}_{\mathbf k}
\delta(\epsilon_{\mathbf k} - \epsilon_F),
\end{multline}
where the scattering rate
$1/\tau({\mathbf k}) = \sum_{{\mathbf k}^\prime} C(\mathbf k,{\mathbf
k}^\prime)$ and
$C(\mathbf k,{\mathbf k}^\prime) = 2\pi \delta(\epsilon_{\mathbf k} -
\epsilon_{{\mathbf k}^\prime}) \tau^{-1}(\theta,\theta^\prime)$.  One
must solve for $g({\mathbf k})$, the deviation from the equilibrium distribution.

Our numerical calculations involve constructing the matrix $A$ of
Eq.~\ref{adef-eqn}, for an arbitrary, discretized Fermi surface and
the scattering rate in Eq.~\ref{mt-thetathetaeqn}; sums are weighted
appropriately by the density of states per unit length along the
surface, and $g({\mathbf k})$ thus obtained directly.

For our analytic calculations (following VA), the Boltzmann equation
is solved by expanding to first order in $\theta_c^2$, in addition to
the Zener--Jones expansion.  The derivatives caused by integrating
with a sharply peaked function turn the scattering time into an
operator, labeled by VA as $\hat\tau$.  Noting that the integral of
$\tau^{-1}(\theta,\theta^\prime)$ over $\theta^\prime$ must yield the
ARPES inverse lifetime (Eq.~\ref{mt-arpesgamma}), the transport
equation (\ref{adef-eqn}) is reduced to a relationship which must be
solved for $\hat \tau(\theta)$.


We illustrate the form of our results with an explicit example,
within which coefficients can be calculated analytically.
For a circular Fermi surface, the equation for $\hat \tau(\theta)$ is:
\begin{equation}
\label{tauhatcirceqn}
\hat\tau v = \tau_M v
+ \frac{\theta_c^2 \tau_M}{\tau_i}
  \frac{\mathrm{d}}{\mathrm{d}\theta}
  \left[
	 \cos^2 2\theta \frac{\mathrm{d}(\hat\tau v)}{\mathrm{d}\theta}
	\right].
\end{equation}
Noting that derivatives of $\hat\tau v$ only enter at the next order
in $\theta_c^2$ means we can write ${(\hat\tau v)}^{\prime} \simeq
\tau_M v^\prime$ etc. and so
\begin{equation}
\hat\tau(\theta) = \tau_M + \frac{\theta_c^2 \tau_M^2}{\tau_i}
\left(
\cos^2 2\theta \frac{\mathrm{d}^2}{{\mathrm{d}\theta}^2}
- 2 \sin 4\theta \frac{\mathrm{d}}{{\mathrm{d}\theta}}
\right).
\end{equation}
This is equivalent to Eq.~16 of Ref.~\onlinecite{va}, except the
dependence on the original scattering times and $\theta_c$ is shown.
Conductivities are obtained from expressions such as
\begin{equation}
\sigma^{xy}
=
\frac{n e^2}{\pi m}
\frac{eB}{m}
\int \mathrm{d}\theta  v^{x}
\hat\tau  \frac{\mathrm{d}}{\mathrm{d}\theta}  \hat\tau  v^y,
\end{equation}
and we find
\begin{align}
\sigma^{xx} &\sim
&\tau_M
&\left(C^0_{xx} - C^1_{xx}\frac{\theta_c^2
\tau_M}{\tau_i}
+O(\theta_c^4) \right),
\label{sigmaxxeqn}
\\
\sigma^{xy} &\sim 
& \omega_c \tau_M^2
&\left(C^0_{xy} - C^1_{xy} \frac{\theta_c^2 \tau_M}{\tau_i}
+O(\theta_c^4) \right),
\label{sigmaxyeqn}
\\
\Delta\sigma^{xx} &\sim
&- \omega_c^2 \tau_M^3
&\left(C^0_{\Delta xx} - C^1_{\Delta xx}\frac{\theta_c^2
\tau_M}{\tau_i}
+O(\theta_c^4) \right).
\label{deltasigmaxxeqn}
\end{align}
This shows the apparent 
\footnote{However, this apparent new temperature dependence 
can be misleading: for example in the case
of small-angle impurity scattering which is {\em isotropic} 
around the Fermi surface these terms can be summed to all orders in
$\theta_c$ to give a temperature independent addition to the resistivity
({\it i.e.} Matthiessen's rule). Since we will be looking for a large
correction at leading order in the analytic work described here this
effect will not be important. In the numerical work described later,
we effectively do work to all orders in $\theta_c$ so take this effect
into account.} appearance of new $T$-dependences at each order in
$\theta_c^2$. The essence of VA's argument rests on the value of the
numerical coefficients $C$. To account for the measured resistivity,
$C_{xx}^1/C_{xx}^0$ must be generally small: small-angle scattering
does not dramatically affect the marginal Fermi-liquid scattering
rate so $\rho_{xx} = \sigma_{xx} \sim \tau_M^{-1} \sim T$. 
However, for the Hall conductivity VA argue that
$C_{xy}^1/C_{xy}^0$ is large so that the first correction dominates
the zeroth order term. This would have a dramatic effect on the
inverse Hall angle, since 
$\cot\Theta_H = \sigma^{xx}/\sigma^{xy} \sim \tau_M^{-2} \sim T^2$,
thus generating an apparently new temperature dependent scattering
rate. 

For the circular Fermi surface we can calculate these coefficients and
find 
\begin{equation}
C^0_{xx,xy,\Delta xx}=1; \quad C^1_{xx}=1, C^1_{xy}=2, C^1_{\Delta xx}=3.
\end{equation}
When substituted into the expression for the inverse Hall angle above
we see that for this case, to order $\theta_c^2$, the inverse Hall
angle and the resistivity have an identical temperature dependence and
there is no new effect from small-angle scattering.

Clearly then, for small-angle scattering to have any effect, further 
anisotropy must be included~\cite{vaerratum}. Also we have assumed that
$\theta_c^2$ is a reasonable expansion parameter but, if the leading
correction is to ever dominate the zeroth order term, this must be
questionable. To address both these issues we now adopt a numerical
approach which makes no assumptions on the size of $\theta_c$ and can
handle arbitrary Fermi surfaces and other anisotropies. 

We have solved the model of Eq.~\ref{mt-thetathetaeqn}
and~\ref{adef-eqn} numerically without recourse to an expansion, so
any new temperature dependence in the Hall angle can be seen
directly. We studied a wide variety of Fermi surfaces and, in
particular, the ARPES best-fit parameterization~\cite{norman1}. We
also allowed the small scattering angle itself to vary around the
Fermi surface [for example as $\theta_c \sim 0.1 (1 + \cos^2
2\theta)$]. The results of this study were that we were unable to find
any significant new temperature dependence appearing in the Hall angle
with the exception of the particle-hole symmetric Fermi surface
discussed below.

To make quantitative comparison with the analytic work on the circular
Fermi surface above we used the numerics to extract the numerical
coefficients of the expansion of
Eqs~\ref{sigmaxxeqn},\ref{sigmaxyeqn},\ref{deltasigmaxxeqn}. We found
that for the Fermi surface of Ref.~\onlinecite{norman1} the ratios
$C_{xx}^1/C_{xx}^{0}$ and $C_{xy}^1/C_{xy}^{0}$ never varied from the
circular Fermi surface values of 1 and 2 respectively by more than
50\%. These values were typical for a range of similar cuprate-like
Fermi surface parameterizations. Our conclusion is that
$C_{xy}^1/C_{xy}^0$ is not, in general, large enough to generate a new
temperature dependence in the Hall angle.

\begin{figure}
\includegraphics[width=0.5\columnwidth]{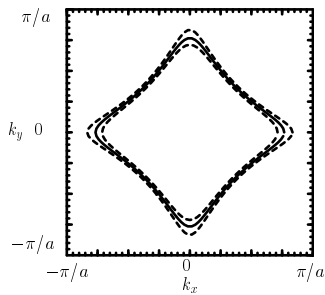}
\caption{\label{fsfig}An example of a ``particle-hole symmetric'' Fermi
surface, for which the conventional Hall term vanishes
($\epsilon_{\mathbf k} = - t(\cos k_x + \cos k_y + 0.5966 \cos k_x
\cos k_y)$; solid line: $\epsilon_F = -0.70t$).  Note that the
shape does not coincide well with observed cuprate Fermi surfaces.  A
linear-$T$ Hall angle has returned when the surface is tuned as far as
either of the dashed shapes ($\epsilon_F = -0.75t, -0.65t$).}
\end{figure}

\begin{figure}
\includegraphics[width=0.9\columnwidth]{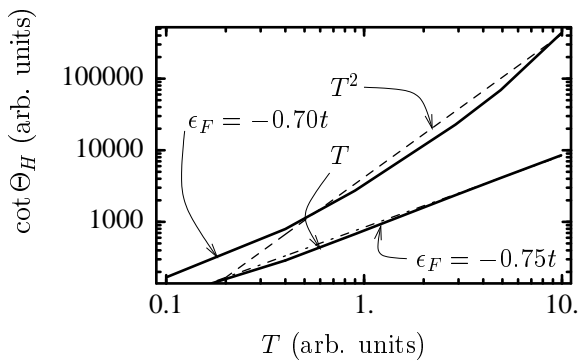}
\caption{\label{cotthfig} Log-log graph of $\cot \Theta_H$ against
temperature (arbitrary units).  A $T^2$ law (dashed line) is possible
for the right Fermi surface (as Fig~\ref{fsfig}, $\epsilon_F =
-0.70t$) but on altering the Fermi level slightly, the linear-$T$ law
(dot-dashed line) quickly returns ($\epsilon_F = -0.75t$).
[Parameters used: $\theta_c \sim 0.02 $, $\tau_i^{-1} \sim 10$,
$\tau_M^{-1} = 0.1 + T$.]}
\end{figure}

The one exception to our findings is for Fermi surfaces very close to
particle-hole symmetry when the Hall conductivity changes sign; at
this point by definition the conventional term $C^0_{xy}$ vanishes.
This corresponds to a ``net curvature'' of zero, and an example is
shown in Fig~\ref{fsfig}.  The new term $C^1_{xy}$ does not vanish
here, and will dominate the Hall effect.  Hence the small-angle
scattering model can reproduce the experimental temperature
dependences (see Fig.~\ref{cotthfig}), but only for specially tuned
Fermi surfaces.

However there are significant differences between this model close to
particle-hole symmetry and what is observed in the cuprates. 
As one moves from
underdoped to overdoped, the coefficient of $T^2$ is observed to
change, but there is only a slight variation in the
exponent~\cite{carrington1,ando1}. Within the small-angle scattering model,
adjusting the Fermi level through a similar range causes significant
violations to the power law.  Moreover, it can only provide a return
toward the intrinsic linear-$T$ behavior (with a sign change in the
Hall angle on one side), in contrast with the continuous small
increase in exponent observed experimentally.  Also, the size of
$1/\omega_c \tau$ decreases by an order of magnitude in the model,
unlike the cuprates where the value of the Hall angle at a given
temperature does not vary so strongly as a function of doping.

Here we are assuming a na{\"\i}ve `rigid band' picture of the doping
dependence of the Fermi surface geometry.  There may be correlation
effects that `pin' this Fermi surface to particle-hole
symmetry,\cite{wang1} so this sensitivity alone does not preclude the
model.  Hence we consider the magnetoresistance.


Recall that the magnetoresistance is a combination of the
magnetoconductance and the Hall angle:
\begin{equation}
\label{mt-deltarhoeqn}
\frac{\Delta\rho^{xx}}{\rho^{xx}}
=-\frac{\Delta\sigma^{xx}}{\sigma^{xx}}
-{\left(\frac{\sigma^{xy}}{\sigma^{xx}}\right)}^2.
\end{equation}
To account for experimental measurements we must have
$\Delta\rho^{xx}/\rho^{xx}\sim \Theta_H^2 \sim {(\sigma^{xy}/\sigma^{xx})}^2$.
The change in the conductivity $\Delta\sigma^{xx}$ due to the Lorentz
force is always negative, and the orbital
magnetoresistance $\Delta\rho/\rho$ is always positive; thus
${(\sigma^{xy}/\sigma^{xx})}^2 <
|\Delta\sigma^{xx}/\sigma^{xx}|$ and the Hall angle \emph{in itself}
cannot be enough to explain magnetoresistance measurements, but will
be a \emph{negative} correction to the first term.

Ong\cite{ong1} related the magnetoresistance to the variance of the
Hall angle around the Fermi surface, and hence it might be argued (as
VA do) that the temperature dependence of the magnetoresistance
follows from that of the Hall angle.  However, notwithstanding the
arguments of the previous paragraph, Ong's proof relies upon a Stokes
integral of mean free path, which apparently cannot be constructed
in the situation where we have both a operator $\hat\tau$ and a
non-constant Fermi speed.

Thus, we emphasize that it is essential to calculate the
magnetoresistance directly, via the magnetoconductance (\textit{e.g.}
Eq.~\ref{deltasigmaxxeqn}).  To generate a $T^{-4}$ temperature
dependence it is required that $\Delta\rho/\rho \sim \tau_M^4$ (since
$\tau_M$ is our only $T$-dependent parameter).  Whilst the
contribution to $\Delta\rho/\rho$ from the Hall angle,
${(\sigma^{xy}/\sigma^{xx})}^2$, will produce this at $O(\theta_c^4)$
for the tuned Fermi surface, it cannot dominate the magnetoresistance
(as we have just argued), and so the only possibility for matching the
experiments comes from an additional $O(\theta_c^4)$ term in the
magnetoconductance $\Delta\sigma^{xx}$ (Eq.~\ref{deltasigmaxxeqn}):
$C^2_{\Delta xx} \theta_c^4 \tau_M^2 / \tau_i^2 $.  This term would
have to dominate.  However, $C^2_{\Delta xx}$ is not large enough, and
the Fermi surface cannot be further fine tuned to make the leading
order terms at $O(\theta_c^0,\theta_c^2)$ disappear: $C^0_{\Delta xx}$
is positive definite and, unlike the corresponding term $C^0_{xy}$ for
the Hall angle, can never vanish.

\begin{figure}
\includegraphics[width=0.9\columnwidth]{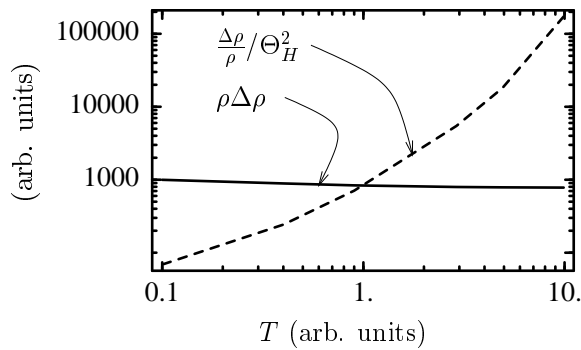}
\caption{\label{kohlerfig}Log-log graph of $\rho\Delta \rho$ (solid
line) and $\frac{\Delta\rho}{\rho}/\Theta_H^2$ (dashed line) against
$T$ (parameters as Fig~\ref{cotthfig}, special Fermi surface
$\epsilon_F = -0.70t$).  Kohler's rule $\Delta\rho/\rho\sim\rho^{-2}$
can be seen to be obeyed ($\rho\Delta \rho \sim T$ independent).
Experiment~\cite{harris1,malinowski1} suggests a ``modified Kohler's
rule'' ($\Delta\rho/(\rho \Theta_H^2) \sim T$ independent) which
cannot be reproduced within the model.}
\end{figure}

This argument is verified by extensive numerical study, in which we
are unable to find a magnetoresistance with much different from a
$T^{-2}$ law, whether or not the Fermi surface is particle-hole
symmetric.  We illustrate our findings with a representative plot
(Fig.~\ref{kohlerfig}), which shows that Kohler's rule is well obeyed,
in contradiction with the experimental measurements.

In conclusion, Varma and Abraham's suggestion that the inverse Hall
angle is measuring the square of a relaxation rate provides an
important new perspective on the unusual magneto-transport properties
of the cuprates. Their prediction of the form of the optical Hall
angle is powerful evidence in favor of their suggestion. In this
paper we have sought to analyze whether a model of small-angle
scattering in a marginal Fermi-liquid can account for the form of Hall
angle that Varma and Abrahams suggest. We have studied such a model
both analytically and numerically, including various anisotropies, and
have also gone beyond the Hall conductivity to consider
magnetoresistance.

We find that in general small-angle scattering does not
discriminate significantly differently between resistivity and
inverse Hall angle except very close to particle-hole symmetry. At
this point however the magnetoresistance continues to be dominated by
the scattering rate seen in the resistivity. This conventional
Kohler behavior of the magnetoresistance is found in this model 
independent of proximity to particle-hole symmetry and in contrast to
the cuprates which show strong deviations from Kohler's rule. Thus we
conclude that small-angle scattering in a marginal Fermi-liquid is not
the origin of the unusual magnetotransport in the cuprates. This
problem remains a tantalizing key to deciphering 
the unconventional normal state properties of the cuprates.

We thank H. D. Drew, M. Grayson, C. Hooley and V. Yakovenko for
helpful discussions. We gratefully acknowledge correspondence with
E. Abrahams and C. M. Varma at whose suggestion we considered the role
of $\theta_c$ anisotropy and Matthiessen's rule physics within this
model. This work was supported by the Royal Society and the Leverhulme
Trust (AJS) and EPSRC (ECC).


\end{document}